\documentclass[final]{svjour3}
\usepackage{graphicx}
\usepackage{rotating}
\usepackage{amsmath,amssymb}
\usepackage{mathptmx}
\usepackage[numbers]{natbib}
\usepackage{mathrsfs}
\makeatletter
\journalname{Journal of Low Temperature Physics}
\bibpunct{}{}{,}{s}{}{,}

\begin{document}

\newcommand{\hdblarrow}{H\makebox[0.9ex][l]{$\downdownarrows$}-}
\title{Application of deep learning to the evaluation of goodness in the waveform processing of transition-edge sensor calorimeters}

\author{Y.\,Ichinohe$^{1}$ \and S.\,Yamada$^{1}$ \and R.\,Hayakawa$^{2}$ \and S.\,Okada$^{3}$ \and T.\,Hashimoto$^{4}$ \and H.\,Tatsuno$^{2}$ \and H.\,Suda$^{2}$ \and T.\,Okumura$^{5}$}

\institute{1. Department of Physics, Rikkyo University, Tokyo 171-8501, Japan\\ \email{ichinohe@rikkyo.ac.jp}\\
2. Department of Physics, Tokyo Metropolitan University, Tokyo 192-0397, Japan\\
3. Engineering Science Laboratory, Chubu University, Aichi 487-8501, Japan\\
4. Japan Atomic Energy Agency (JAEA), Ibaraki 319-1184, Japan\\
5. Atomic, Molecular and Optical Physics Laboratory, RIKEN, Saitama 351-0198, Japan
}

\maketitle

\begin{abstract}
Optimal filtering is the crucial technique for the data analysis of transition-edge-sensor (TES) calorimeters to achieve their state-of-the-art energy resolutions. Filtering out the `bad' data from the dataset is important because it otherwise leads to the degradation of energy resolutions, while it is not a trivial task. We propose a neural network-based technique for the automatic goodness tagging of TES pulses, which is fast and automatic and does not require bad data for training.


\end{abstract}

\section{Introduction}
Transition-edge-sensor (TES) calorimeters are the key instrument in the next-generation high-resolution X-ray spectroscopy. In order to achieve their state-of-the-art energy resolutions, the optimal filtering method is usually used for the energy reconstruction of the data obtained with TES. Mathematically, optimal filtering is a procedure to derive the pulse height $H$, when the input pulse is expressed as $D(\omega)=HS(\omega)+N(\omega)$ in the frequency ($\omega$) domain, where the power spectra of the normalized noiseless signal and noise are $|S(\omega)|^2$ and $|N(\omega)|^2$, respectively.

\begin{figure}[htbp]
\centering
\includegraphics[width=0.6\linewidth, keepaspectratio]{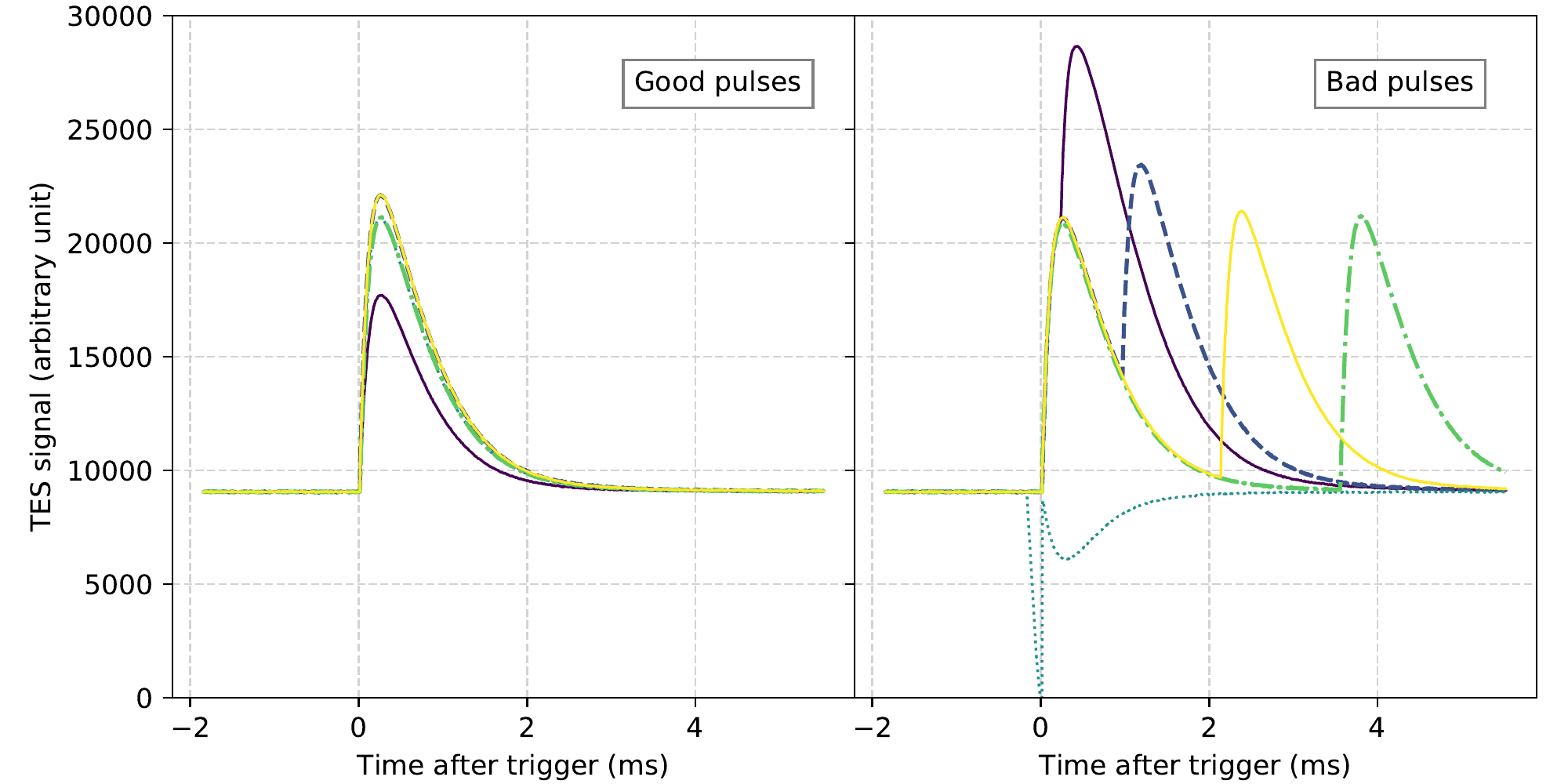}
\caption{Examples of good pulses ({\it Left}) and bad pulses ({\it Right}).}
\label{fig:goodbad}
\end{figure}

As the analytic form of the noiseless signal of a given TES calorimeter $S(\omega)$ is usually unknown, the average pulse calculated in the obtained data itself is often used instead. By using the average pulse, we expect that it represents $S(\omega)$, the ideal pulse shape, as much as possible. Therefore, the existence of the pulses whose shapes significantly deviate from the ideal shape in the dataset from which the average is taken results in the degradation of TES energy resolution. Therefore, it is important to filter out such `bad' pulses (Fig.~\ref{fig:goodbad}).

However, estimating the goodness of individual pulses is not a trivial task. For example, it might be possible to parametrize the pulse shape using e.g. the fluctuation of the baseline level, the slope of the rising edge, and the smoothness of the falling slope. Although these parameters should reflect the behavior of pulses to some extent, such simple parametrizations have several shortcomings; e.g. parameter tuning requires several iterations of trial-and-error processes, leading to a significant amount of overhead in the preprocessing of signal analysis.

Modeling the waveforms by some analytic function and evaluating the residuals by e.g. chi-squared statistics are also possible. However, as the real waveforms are governed by the detector's complex heat, electrical and mechanical responses, they are rarely expressed by simple analytic formulae, and the distribution of such residual statistics should also be complex. Analyzing the complex residual distribution is as hard as analyzing the raw pulses.

Even when such a method is well established, there is another problem; as these methods are basically based on the comparison of some statistical properties between the good and bad pulses, it is quite uncertain that the method actually works on unknown badness --  `bad' examples that do not exist in the training data. Therefore, a goodness evaluation method that is fast and automatic and works for unknown badness is in great demand for the coming era of high-resolution X-ray spectroscopy.

We propose a neural network-based technique for the automatic goodness tagging of TES pulses, which is fast and automatic and does not require bad samples for training. In the following sections, we present the details of our method, demonstrate the performance of the method by actually implementing it, and discuss the result with future prospects.

\section{Methods}

In this work, we employed the neural-network architecture, variational auto-encoder (VAE\cite{vae}) to evaluate the goodness of each waveform. Elaborate explanations of such techniques and the detailed explanation of VAE are presented elsewhere, and the references therein \cite{ichinohe19, lecun15, goodfellow16}.

VAE is a variant of another neural network architecture, auto-encoder (AE\cite{ae}). An AE network consists of an input encoder network and the decoder network that is serially connected just after the encoder network. The output dimension of the encoder network is usually smaller than the input dimension, and the decoder network is trained to reproduce the input data using the small-dimension latent expression. AEs are thus often used to extract the compressed expression of input data, similarly to e.g. the principal component analysis (PCA), but with the high-expressive power realized by the deep non-linear neural network architecture\footnote{In the ideal case where all the pulses are truly identical except for scaling and truly random noises, the first component of PCA is expected to well represent the ideal pulse. However, in the real-world situation, those are not satisfied; pulse shapes may vary non-linearly depending on energy, and noises are sometimes coherent due to cross-talks. It is impossible in principle to cover such variations using PCA because it is basically based on linear transformation. VAE tries to compress such variations without the constraint of linearity; that is, it is possible that a certain latent axis represents the non-linear energy dependence pulse heights. We think this expressive power to be the major advantage of VAE over PCA and other linear methods for the current problem setup.}.

Although VAE is designed to perform the variational Bayesian method to infer the generative model of the training dataset and its philosophy is somewhat different from the ordinary AE, technically, these two architectures are similar. The main difference between the ordinary AEs and the VAEs are as follows; (1) in VAEs, the latent expression is generated probabilistically through random sampling, while it is calculated deterministically in ordinary AEs. (2) in the loss function of VAEs, an additional regularization term is included in addition to the loss function of AEs. The regularization term corresponds to the Kullback-Leibler divergence related to the neural-network approximation of the true generative model of the given dataset (i.e., the first term of Eq. (1) in Ichinohe et al. 2019\cite{ichinohe19}).

When a VAE network is successfully trained using a given dataset, its encoder should have learned to compute the compressed expression of the input vector, while the decoder learned to generate the vector from the compressed expression through the generative model of the training dataset.

Therefore, when inputs are fed into the trained VAE network, the resulting outputs would be twofold; (1) if the input instance is similar to the ones in the training dataset, the corresponding output would be similar to the input because the network is trained so. On the other hand, (2) if the instance is significantly different from the typical instances of the training dataset, the output would be significantly different from the input because the generator has the ability only to reproduce the instances that the generative model of the input dataset would generate. By quantifying the difference of the input and its corresponding output, the similarity of the input to the training data can be evaluated. Thus, if the training dataset consists only of good data, the goodness of the input can be inferred.

\section{Implementation}

\begin{figure}[htbp]
\centering
\includegraphics[width=1.0\linewidth, keepaspectratio]{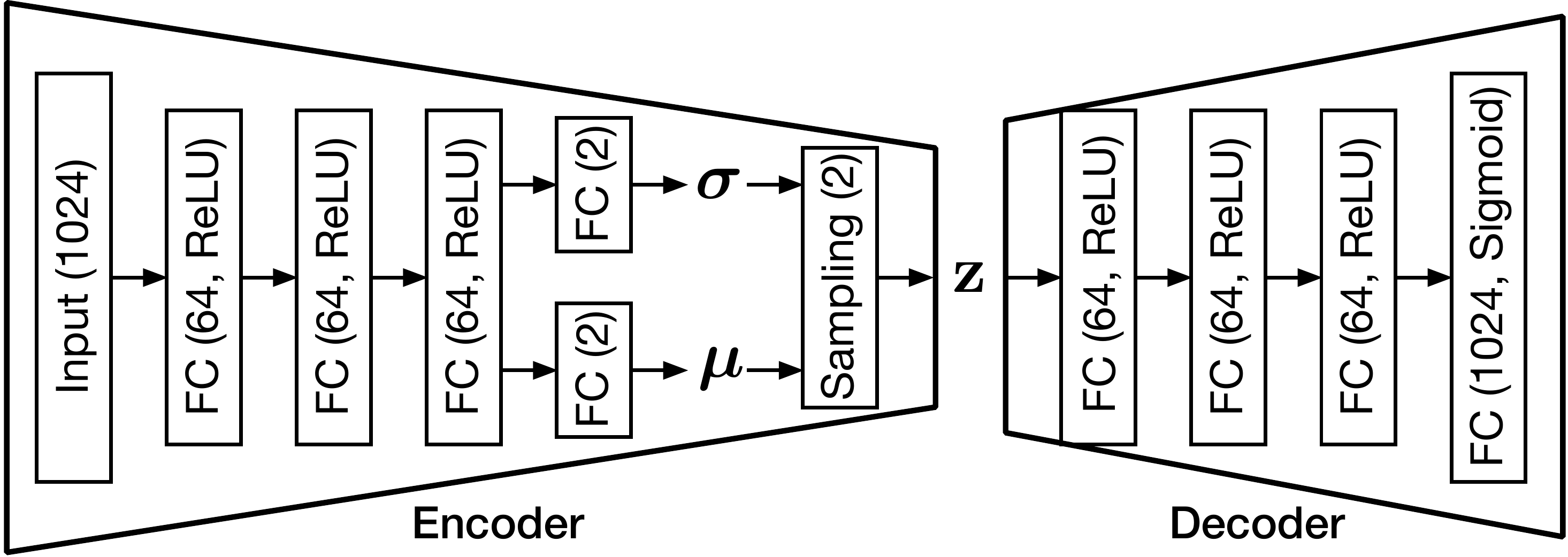}
\caption{Schematic view of the neural network architecture. Three fully-connected (FC) layers are used both in the encoder and decoder. A two-dimensional latent vector $\bf{z}$ is randomly sampled from a multivariate Gaussian distribution using $\bf{z}\sim\mathscr{N}(\bf{\mu},\bf{\sigma}^2\bf{I})$, where $\bf{\mu}$ and $\bf{\sigma}^2\bf{I}$ are the mean and the (diagonal) covariance matrix, which are computed in the last fully-connected layeres of the encoder. The number of the units in each layer and the employed activation function (when applicable) are shown in the parentheses.}
\label{fig:network}
\end{figure}

We implemented the VAE network using the \verb+python+ neural network library \verb+Tensorflow+ 2.3.1\cite{tensorflow}. We employed the simplest, fully-connected architecture for both encoder and decoder. Each fully-connected layer has 64 neurons, and the output is non-linearly transformed using the ReLU\cite{relu} activation function. Only the last layer of the decoder is activated using the sigmoid function so that the output values are in the range between 0 and 1. We set the latent dimension to 2 because the good waveforms are basically similar, and the intrinsic degree of freedom to express the variations of good waveforms should not be large. The conceptual diagram of the network is shown in Fig.~\ref{fig:network}. Although hyperparameter tuning is a essential process in the real-world application, for the present paper, we put an emphasis on demonstrating the idea using a simple setup, and the thorough parameter tuning is left for future works. The best parameters depend on the problem and dataset, as well as how the output of VAE is used in the following processes (i.e., MSE threshold cut in this case).

This study's data acquisition is performed in the hard X-ray experiment campaign at the SPring-8 synchrotron X-ray beamline. A 240-pixel TES spectrometer array developed by NIST\cite{doriese09} was irradiated by a $^{55}$Fe radioactive source under several beam conditions. 
For the baseline `clean' dataset that is used for the VAE training, data collected while the beam was off are used. For performance evaluation, two datasets -- the datasets taken when the TES array trigger rate was relatively low (low-rate dataset) and high (high-rate dataset) were used.

The record length of each pulse is 1024, corresponding to the real time of $\sim$7.4~ms. The total pulses in the baseline, low-rate, and high-rate datasets are 11213, 1969, and 21075, respectively. All the pulses are preprocessed using the min-max normalization using the individual minimum and maximum values.

We trained the VAE network using the baseline `clean' dataset. The training was performed over 1000~epochs with the Adam algorithm\cite{adam} as the network optimizer. The time taken for the network training was $\sim$10~min with the batch size of 64, using the commercial graphics card, NVIDIA GeForce GTX 1080 Ti. The prediction using the network is extremely rapid because it is optimized for the use of GPUs. For example, 10000 waveforms can be processed in less than one second with the present network configuration and computational environment.

\section{Results and discussion}

\begin{figure}[htbp]
\centering
\includegraphics[width=1.0\linewidth,keepaspectratio]{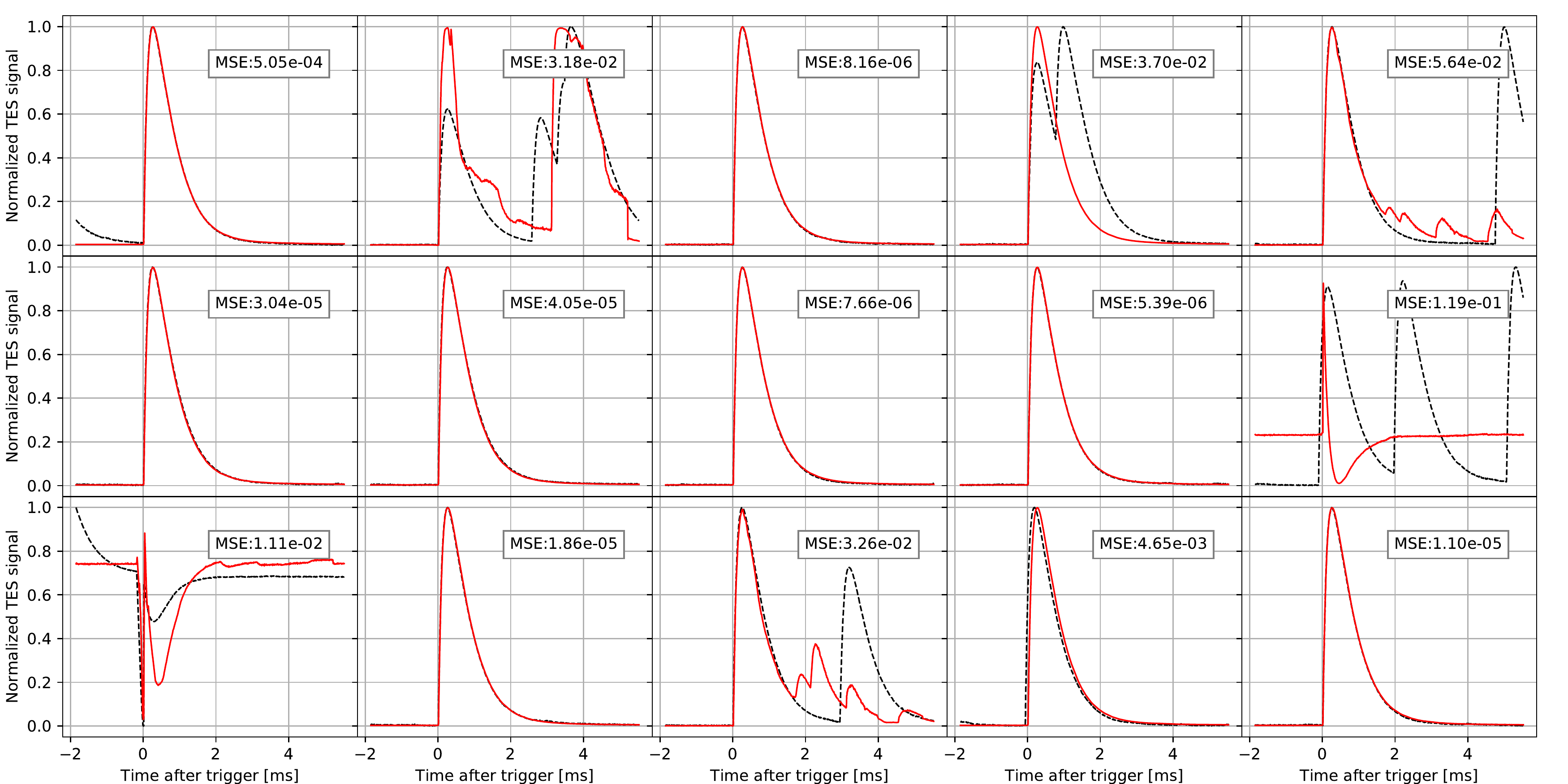}
\caption{Raw waveforms (black dashed lines) and the corresponding VAE-reconstructed waveforms (red solid lines). The corresponding MSE value (see the main text) is also shown in each panel.}
\label{fig:diff}
\end{figure}

Fig.~\ref{fig:diff} shows examples of the network behavior for a variety of input waveforms. The black and red waveforms are the inputs and the corresponding outputs, respectively. When the input waveform is `good', the corresponding reconstructed waveform is almost identical to the original input. When the input waveform is `bad', the corresponding output is either very distorted (e.g. top-row, rightmost) or similar to the `good' waveforms (e.g. top-row, leftmost), resulting in the apparent difference between the input and output in both cases.

To quantify the difference of the input and the corresponding reconstructed output, we calculated the mean-squared error (MSE) value; MSE$=<(X_\mathrm{pred}-X_\mathrm{orig})^2>$, where $X_\mathrm{orig}$ and $X_\mathrm{pred}$ are the (min-max-normalized) input and output pulses, respectively. The MSE values are also presented in the insets of Fig.~\ref{fig:diff}. As expected, the MSE values are smaller for `good' inputs because of similar input and output and conversely larger for `bad' inputs due to the difference of input and output. Note that there is a possibility that the VAE accidentally reconstruct `bad' pulses (see, e.g., the panel at the top row of the second to left column, in which two pulse-like shapes appear in the reconstructed waveform). However, it does not practically matter whether or not it reconstructs some fraction of bad examples. As we do not know the true good pulse, we cannot evaluate the methods based on specific instances. What we can do instead is to evaluate the resulting products, such as the energy resolution or the average pulse.

\begin{figure}[htbp]
\centering
\includegraphics[width=0.5\linewidth, keepaspectratio]{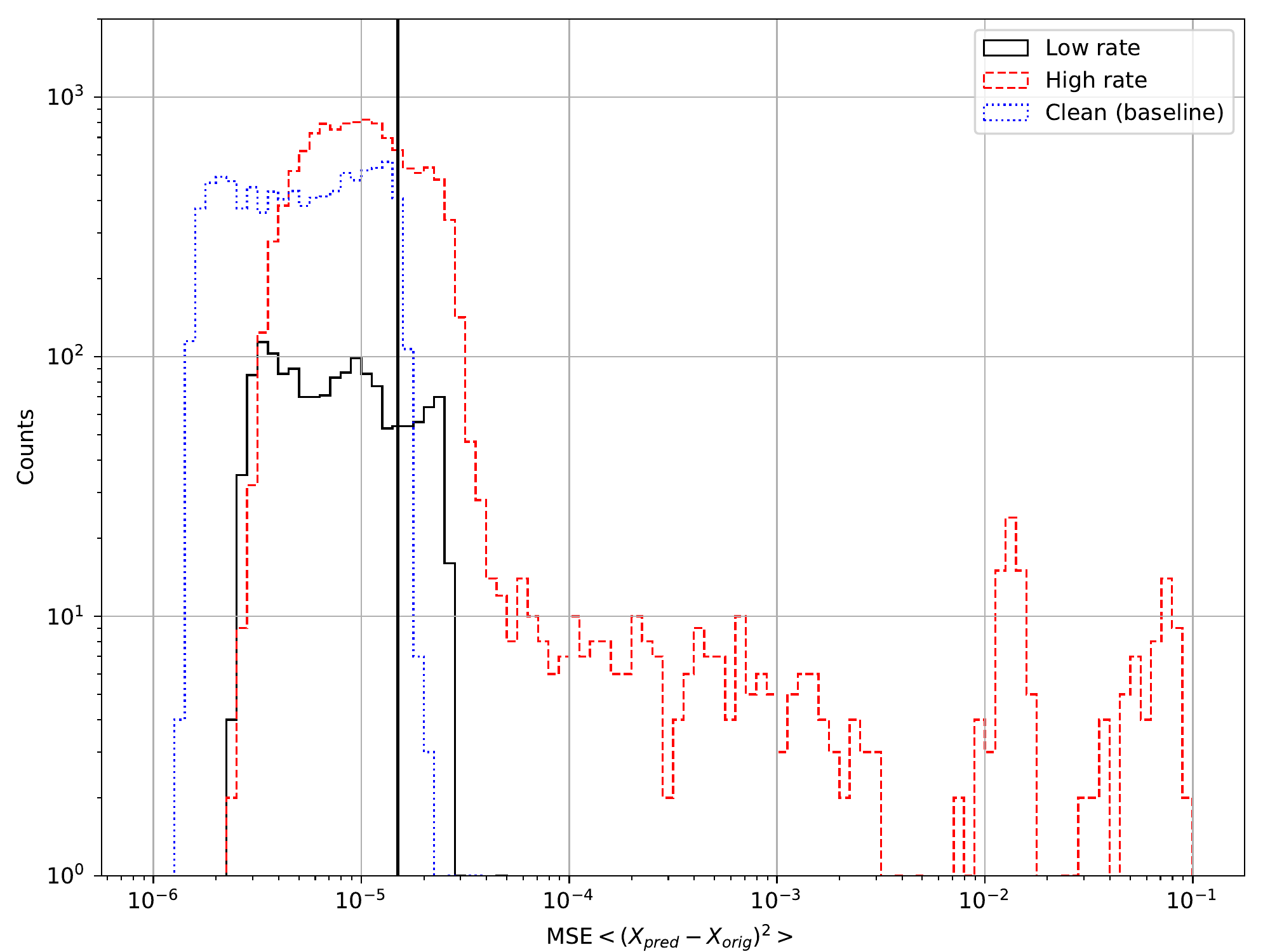}
\caption{The distributions of the statistic MSE$=<(X_\mathrm{pred}-X_\mathrm{orig})^2>$ for the low-rate dataset (black solid line), the high-rate dataset (red dashed line) and the baseline clean dataset (blue dotted line). The black vertical line at $MSE=1.5\times10^{-5}$ denotes the example goodness threshold quoted in the main text.}
\label{fig:vaecomp}
\end{figure}

Fig.~\ref{fig:vaecomp} shows the distributions of the MSE value for the `clean' (blue dotted line), `low-rate' (black solid line), and `high-rate' (red dashed line) datasets. We find that the number of high-value instances increases as the dataset becomes less clean, as expected. For demonstration purposes, we set the goodness threshold of the MSE value as 1.5$\times10^{-5}$ (black vertical line in Fig.~\ref{fig:vaecomp}), based on the observation that most instances in the clean dataset are classified as `good' with this value. Note that the choice of the threshold is somewhat arbitrary, and the best value depends on the situation; the higher (looser) MSE threshold results in less discrimination power but saves more pulses. Conversely, the lower (stricter) one leads to the cleaner filtered data, but the number of the remaining instances is smaller.

\begin{figure}[htbp]
\centering
\includegraphics[width=1.0\linewidth,keepaspectratio]{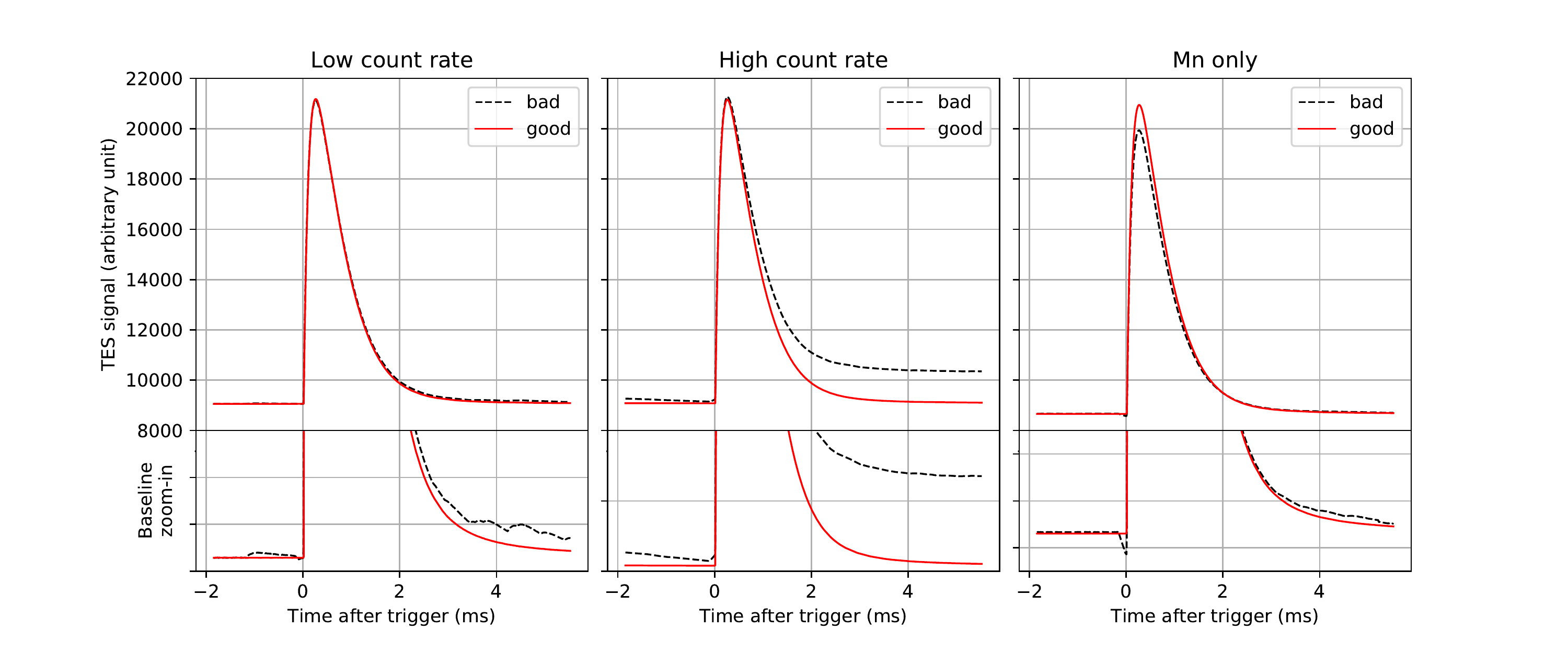}
\caption{The comparison of the average pulse computed using only the good pulses (red solid lines) and the bad pulses (black dashed lines), with the goodness threshold of the MSE value of 1.5$\times10^{-5}$. The corresponding zoom-in to the baseline level is shown below. The low-rate, high-rate, and the baseline clean datasets correspond to the left, middle, and the right panels, respectively.}
\label{fig:result}
\end{figure}

For comparison, we implemented a simple, non-parametric modeling method. We prepared the template waveform by simply taking average of the waveforms in the training dataset. Each waveform in the evaluation datasets are fitted by the template, and the MSE value between the scaled template and the input pulse are analyzed as in the last paragraph.
We found that the true positive rate (performance of detecting `bad' pulses) derived using VAE is better than that using the simple fitting method for a similar true negative rate (saving `good' pulses).
We think that this is mainly because VAE can catch small anomalies such as accidental crosstalks from the surrounding pixels. The absolute intensities of such anomalies are small, and fitting-based methods are designed to reduce MSE. This combination renders the prominence of small anomalies weaker, leading to worse performance on catching such deviations. On the other hand, VAE is designed to generate the `ideal' pulse corresponding to the input waveform, where unwanted tweaks such as baseline-level fitting are avoided.

Fig.~\ref{fig:result} compares the average pulse computed using only the good pulses and the bad pulses with the goodness threshold defined in the previous paragraph. We found that the good average pulses behave well for all the cases, and bad average pulses are apparently `bad'. This indicates that with our proposed method, the good and bad pulses are automatically discriminated, and thus demonstrates the effectiveness of the method.

It is worth pointing out that the bad average pulse of the `clean' dataset shows apparent deviations from the corresponding good average pulse. This indicates that the rare, small number of `bad' instances in the training dataset (which is expected to be mostly `good') can also be automatically extracted. The possible reason is that bad instances are not typical in the `clean' dataset, and it is more difficult for the VAE network to reconstruct such rare instances properly, resulting in slightly higher MSE values of such bad instances residing in the `good' dataset. Although this method should work ideally if the training dataset contains only good pulses (no bad pulses at all), we think that the method is relatively insensitive to the existence of some `bad' samples in the training data.

\section{Summary and future prospects}

We proposed a fast, neural network-based method for automatic goodness evaluation of the waveforms obtained by TES calorimeters and demonstrated the concept using real TES data. Unlike typical classification models, the method does not require `bad' data for training. This is a significant advantage of our model because it is often difficult to collect enough numbers of `bad' instances because most instruments are basically designed to obtain `good' data, and it is known that imbalanced training datasets often result in poor performance of the trained model.

Considering that TES pulses represent the instrument's electrical and thermal response, we expect that the same parameters work, at least for many different datasets obtained by the same instrument. Therefore, we think it is safe to use the same network without any tuning once the hyperparameter tuning is done.

There are sometimes cases that it is difficult even to formulate the badness of the signal. For example, when the detector response is very complex, the analytic expression of the data instance might be intractable, prohibiting the modeling-based methods. Our method does not depend on prior knowledge about the data and thus works in such a situation.

Our proposed method is rather data-driven and does not depend on the details of the data. Hence, the method can be applied to any data format other than TES waveforms. As the grading of the data is an essential process in experimental physics, we think the method is widely applicable to topics in a variety of research fields.

Fitting-based and other complex signal filtering methods are mostly designed for offline analysis with sufficient resources. However, it is often necessary to perform the signal filtering very quickly in a limited resource condition in the real-world TES operation, such as a satellite mission, in which it is neither possible to store nor to downlink all the data. Neural-network models in general consist of multiple simple operations and so it is easy to implement such models in the readout digital electronics (e.g. FPGA) directly. Therefore, our proposed method would maximize the capability of future missions with which a huge amount of data are expected to be obtained.

\begin{acknowledgements}
We thank the referees for their constructive suggestions and comments. This work was partialy supported by the Grants-in-Aid for Scientific Research by the Japan Society for the Promotion of Science with KAKENHI grant Nos 18H05458, 20K14524, and 20K20527.
\end{acknowledgements}

\section*{Data availability}
The datasets generated during and/or analysed during the current study are available from the corresponding author on reasonable request.

\pagebreak

\end{document}